\documentclass[preprint,aps,prl,showpacs]{revtex4}
\usepackage[]{epsfig}

\newcommand{\CaW}{\mbox{$\mbox{CaWO}_4$}}
\newcommand{\CoS}{\mbox{${}^{60}\mbox{Co}$}}
\newcommand{\Fe}{\mbox{${}^{55}\mbox{Fe}$}}
\newcommand{\SiN}{\mbox{$\mbox{SiN}_x$}}
\newcommand{\micr}{\mbox{$\mu\mbox{m}$}}
\newcommand{\micrs}{\mbox{$\mu\mbox{s}$}}
\newcommand{\Kf}{\mbox{${}^{40} \mbox{K}$}}
\newcommand{\sys}{\stackrel{sys}{\pm}}
\newcommand{\MPP}{Max-Planck-Institut f\"ur Physik, F\"ohringer Ring 6, D-80805 Munich, Germany}

\begin{document}

\title{A Textured Silicon Calorimetric Light Detector}

\author{P.~C.~F.~\surname{Di Stefano}\footnote{E-mail: distefano@ipnl.in2p3.fr, Permanent address : Institut de Physique Nucl\'eaire de Lyon, 4 rue Enrico Fermi, F-69622 Villeurbanne Cedex, France}}
\affiliation{\MPP}
\author{T.~Frank}
\affiliation{\MPP}
\author{G.~Angloher}
\affiliation{\MPP}
\author{M.~Bruckmayer}
\affiliation{\MPP}
\author{C.~Cozzini}
\affiliation{\MPP}
\author{D.~Hauff}
\affiliation{\MPP}
\author{F.~Pr\"obst}
\affiliation{\MPP}
\author{S.~Rutzinger}
\affiliation{\MPP}
\author{W.~Seidel}
\affiliation{\MPP}
\author{L.~Stodolsky}
\affiliation{\MPP}
\author{J.~Schmidt}
\affiliation{Institut f\"ur Solarenergieforschung Hameln/Emmerthal, Am Ohrberg 1, D-31860 Emmerthal, Germany}

\date{\today}

\begin{abstract}
We apply the standard photovoltaic technique of texturing to reduce the reflectivity of
silicon cryogenic calorimetric light detectors.  
In the case of photons with random incidence angles, 
absorption is compatible with the increase in surface area. For the geometrically thin detectors studied, energy resolution from athermal phonons, 
dominated by position dependence,
is proportional to the  surface-to-volume ratio.  With the \CaW\ scintillating crystal used as light source, the time constants of the calorimeter should be adapted to the relatively slow light-emission times.
\end{abstract}


\maketitle

\section{Introduction}

Cryogenic calorimeters, in which  the  phonons  created 
by incoming particles are read out, now rival longer-established techniques of particle
detection such as ionization in semiconductors and scintillation.  
They boast excellent thresholds and resolutions which can be enhanced by measuring athermal phonons in addition to the thermal ones.
Another of their advantages, exploited by rare-event searches for which radioactive background is an issue, is the
ability to distinguish between particles interacting with electrons in
matter (e.g. photons and electrons) and those interacting with nuclei 
(e.g. neutrons and putative dark matter particles).  
Until now this has been
achieved mainly through a simultaneous measurement of charge in
semiconducting calorimeters~\cite{art:Spooner1991,art:Shutt,proc:Edelweiss2}.  
Another technique is a simultaneous
measurement of scintillation, with a principal calorimeter  made out of a scintillating material which emits photons read in a
secondary calorimeter~\cite{art:Bobin_Scint1997,art:Meunier1999} (or some other light sensitive device~\cite{art:Allesandrello_Scint1998}). 
For instance, the next phase of the CRESST (Cryogenic Rare Event Search with Superconducting Thermometers) dark-matter search will deploy up to 33 such modules with \CaW\ as the main calorimeter~\cite{preprint:CRESST_proposal_2001}.
The challenge is that the emitted light is but a
small fraction of the deposited energy, and not all of it
necessarily reaches the secondary calorimeter.  We report on optimization of
these light detectors.

In the case of main calorimeters like \CaW\  emitting light in the visible
spectrum, 
and for optical sources in general,
silicon would appear well suited as an absorber for
the light-detection calorimeter, because of its band gap around 1~\micr\ (1.17~eV at mK
temperatures).
Moreover, Si has already been successfully used
as an absorber in cryogenic calorimeters (e.g. Ref.~\cite{art:proebst_1995}).  
Its advantages include 
a high speed of sound 
($\approx 5760 \mbox{ m/s}$)
which gives good phonon properties, 
and a
high melting point ($\approx 1690 \mbox{ K}$) which facilitates deposition
of thin films made from materials with high melting temperatures, such 
as tungsten, when they are chosen as thermometers.
However, polished silicon has a high visible reflectivity.  
A similar problem has been encountered in the field of photovoltaics, and
solved by a combination of texturing the surface of the silicon and coating
it with anti-reflective layers~\cite{proc:arndt1975}.  
The texturing squares the reflectivity for normal incident photons by providing 
them with two chances to be absorbed.  We first describe preparation of our
textured light detectors before discussing experimental results obtained.

\section{Preparation of the light detectors}
Two 4 inch diameter, $525 \pm 35 \ \micr$ thick, float-zone p-type
silicon wafers with a resistivity of between 10200 and
71030 $\Omega \mbox{m}$ were used.
Orientation of both wafers was (100) for
the purpose of texturing. One side of each wafer was polished, the
other lapped and etched. A natural silicon oxide layer is assumed
to have been present on all Si surfaces. A 150~nm thick \SiN\ layer was
deposited by plasma-enhanced chemical vapor deposition through an
Al mask into 5~mm diameter disks on the polished surface of one of
the wafers. This wafer was then etched in a
KOH-isopropanol mix at $75{\ }^{\circ}\mbox{C}$ in order to texture the exposed Si into a
random pyramid structure~\cite{proc:IPAetching1973}.  Typical height of the pyramids
is 2--5~\micr, while the pyramid angle of $70.5^{\circ}$ given by the
crystalline structure of Si means that the textured surface area
is about 1.74 times greater than the original, planar, surface area. The
\SiN\ remained unaffected by the texturing.

Samples of size $20 \times 20 \mbox{ mm}^2$ and  $30 \times 30 \mbox{ mm}^2$
were cut from both wafers.  
Tungsten transition-edge sensors of the 
type depicted in Figure~\ref{distefano_fig:thermometer} were next
deposited onto the samples using a standard procedure developped by the CRESST collaboration~\cite{art:Colling1995} : tungsten
films about 300~nm thick were evaporated at  $550{\ }^{\circ} \mbox{C}$ under $10^{-10} \mbox{ mbar}$
onto the samples; the W was structured by photolithography and a 
$\mbox{KH}_2\mbox{PO}_4$ - $\mbox{KOH}$ - $\mbox{K}_3\mbox{Fe(CN)}_6$ - $\mbox{H}_2\mbox{0}$ solution
to sizes of $2 \times 2 \mbox{ mm}^2$ or $2 \times 3 \mbox{ mm}^2$. 
Electrical contact pads made of 200~nm thick aluminum were then sputtered onto
the tungsten for the readout, as was a 200~nm thick gold thermal 
contact.  Similar aluminum pads were deposited as contacts for a film 
heater used to stabilize the operating temperature of the thermometer 
and to send periodic heat pulses to monitor the stability of the 
detector response. Gold or 
aluminum wires of 25~\micr\ diameter were ultrasonically bonded to the 
pads to provide the thermal or electrical links. The W films were
placed near the center of the Si absorbers, and on the \SiN\ in the
case of the textured samples.  
On silicon, the tungsten reliably gave superconducting
transitions near 20~mK, once it was realized this transition appears to
depend on the natural oxide on which the W is deposited :  when the W was
evaporated onto samples which had been etched in HF just before 
mounting in the deposition chamber,
the transition temperature was of the order of 1~K; when the time lapse
between HF bath and W deposition was of the order of a week the transition
was at about 60~mK.  This is presumably linked to some chemical interaction
between Si and W which is inhibited by the presence of natural oxide.
Such interactions also appear to have been blocked by the \SiN\ layer in the case of the textured absorber, as a transition temperature near 20~mK was obtained.

Three detectors were selected for testing :  $20 \times 20 \mbox{ mm}^2$ and $30 \times 30 \mbox{ mm}^2$ planar Si absorbers, and a $20 \times 20
\mbox{ mm}^2$ textured Si one.  Their characteristics are summarized in
Table~\ref{distefano_tab:si_description}.

\section{Experimental setup and results}

\subsection{Preliminary tests and setup}
All three light detectors were first cooled in a
standard copper holder inside a dilution fridge and exposed 
to a collimated \Fe\ source (5.9~keV photons) to estimate
their intrinsic energy resolution.  
Resolution, 
estimated as the full width at half the maximum (FWHM) of the  5.9~keV  line, 
was  350~eV for the textured detector and 180~eV for the smooth ones.
 However, detector responses in terms of pulse height varied with the position of the collimated spot.  It is quite likely that these resolutions, especially that of the textured detector, contain a contribution from the finite size of the collimated hole.

Next, the three detectors were each placed in a setup inside the fridge to measure their light
absorption.  The setup (Figure~\ref{distefano_fig:setup}), described in detail 
elsewhere~\cite{proc:TorstenComo2001,thesis:Torsten2002}, consisted of a
\CaW\ scintillating crystal of cylindrical shape (35~mm high with a 40~mm
diameter) placed in a concentric 50~mm diameter light collector lined with
a polymer reflective foil~\cite{art:PolyFoil}.  The \CaW\ crystal had 
a non-functioning $5 \times 6 \mbox{ mm}^2$ W film on it. Both ends of the
light collector were lined with the same foil; however one of the ends had
four Teflon pegs to hold the light detector.  In this manner, both sides
of the light detector should have been exposed to any available scintillating
light.  Care was taken to minimize thermal leaks between the calorimeters
and their environment while avoiding spurious light traps in the setup.  To
provide an absolute energy reference, a \Fe\ source illuminated the light
detector from outside the light collector foil, through a  hole
in the light collector's Cu structure behind the light detector 
(for mechanical reasons, the whole large light detector was exposed to the source,  
whereas the small detectors where illuminated through a 14~mm diameter hole).  
An external \CoS\ source (main photon lines
at 1.17~MeV and 1.33~MeV) was used to stimulate scintillating light
from the \CaW\ 
(peak of emission $\approx 440 \mbox{ nm}$, $\mbox{FWHM} \approx 100 \mbox{ nm}$~\cite{art:Treadaway1974}).

The detectors were operated in their superconducting transition by 
stabilizing their baseline temperature through the film heaters.  
This proved a challenge at ground level because the high rate of cosmic-ray-induced background 
interacting in the 266~g scintillator led to pile-up in the light detector, 
especially in the large one.  
Pile-up, a nuisance in itself, 
can also degrade the temperature stabilization of transition-edge sensors.  
The large device was therefore operated with active thermal feedback~\cite{proc:OllieLTD8} 
to shorten pulse times in some runs.  Stability of the small detectors was monitored with a pulser.

\subsection{Pulse shapes in light detector}
Two classes of light detector events were recognizable from their time 
constants (Fig.~\ref{fig:distefano_pulses}).
Fast pulses were caused by direct hits in the light detector (mainly due 
to the \Fe\ source). Slower pulses were events of scintillating light  
(due to interaction in the \CaW\ of the cosmic background and \CoS\, when present).  
That the slow pulses originate in the scintillator was previously 
verified by the coincidences between the light detector and an instrumented 
\CaW\ calorimeter.

The direct events in the detectors 
have been fitted with a model assuming an exponential 
rise (collection and thermalization of athermal phonons in the thermometer), a fast exponential decay (relaxation of thermal phonons in the thermometer through its heat sink, referred to as the athermal signal), and a slower  exponential decay (relaxation of the entire detector, referred to as the thermal signal)~\cite{art:proebst_1995}: 
\begin{equation}
R(t) = H(t) \left[ A_{ath} \left( {\mbox{e}}^{-t/{\tau_{ath}}} - {\mbox{e}}^{-t/{\tau_{rise}}} \right)
	+  A_{th} \left( {\mbox{e}}^{-t/{\tau_{th}}} - {\mbox{e}}^{-t/{\tau_{rise}}} \right) \right]
\label{distefano:eqdirect}
\end{equation}
where $H(t<0)=0$ and $H(t \geq 0) = 1$ and, to simplify, the pulses are assumed to start at $t=0$.
Direct 5.9~keV hits gave typical rise times of $\tau_{rise} \approx 60~\micrs$ with a FWHM 
of 15~\micrs\ for the distribution of these events. 
Both the athermal ($\tau_{ath} \approx 0.7 \mbox{ ms}$) 
and thermal ($\tau_{th} \approx 6 \mbox{ ms}$) components were clearly present, 
with the athermal component making up between 40 and 
90~\% of the total pulse amplitude.  All these parameters 
depended on the detector and to a certain extent the operating temperature.

Scintillation-induced hits with similar energy deposits in the 
light detector had typical rise times of 250~\micrs\ with a FWHM of 
50~\micrs\ for the distribution.  
This slow rise time is interpreted as indicative of a relatively slow scintillation component in \CaW.
The athermal component of the scintillating  pulses was suppressed. 
Thus the light detectors appear to have been too fast for the scintillator, 
and were not quite in a calorimetric mode where they would fully integrate 
the energy of the individual photons emitted by an event in the scintillator~\cite{art:proebst_1995}.  
Light detector response was therefore not optimal. 
To understand the scintillation pulse shapes, 
an exponential decay with time constant $\tau_{scint}$ 
of the light emitted from the \CaW\ scintillator has been assumed :
\begin{equation}
E(t) = H(t) \left[ \frac{{\mbox{e}}^{-t/{\tau_{scint}}}}{\tau_{scint}} \right]
\label{distefano:eqemis}
\end{equation}
In this form, $\int_{0}^{\infty} E(t) \ dt$, the total energy emitted in scintillation, is normalized to unity whatever the time constant.
Convolution with the response of the light detector for direct hits yields the expected shape of the scintillation-induced pulses :
\begin{eqnarray}
S(t) & = & \int_{0}^{\infty} E(u) \ R(t-u) \ du \nonumber \\ 
 & = & H(t) \left[ A_{ath} \left\{ \frac{\tau_{ath}}{\tau_{scint}-\tau_{ath}} \left( {\mbox{e}}^{-t/{\tau_{scint}}} - {\mbox{e}}^{-t/{\tau_{ath}}} \right)
- \frac{\tau_{rise}}{\tau_{scint}-\tau_{rise}} \left( {\mbox{e}}^{-t/{\tau_{scint}}} - {\mbox{e}}^{-t/{\tau_{rise}}} \right) \right\} \right] \nonumber 
\\ 
 &  & \mbox{} + H(t) \left[ A_{th} \left\{ \frac{\tau_{th}}{\tau_{scint}-\tau_{th}} \left( {\mbox{e}}^{-t/{\tau_{scint}}} - {\mbox{e}}^{-t/{\tau_{th}}} \right)
- \frac{\tau_{rise}}{\tau_{scint}-\tau_{rise}} \left( {\mbox{e}}^{-t/{\tau_{scint}}} - {\mbox{e}}^{-t/{\tau_{rise}}} \right) \right\} \right]
\label{distefano:eqscint}
\end{eqnarray}
Should the scintillation time constant be equal to one of the phonon time constants $\tau$, the relevant term would become $\frac{t}{\tau} {\mbox{e}}^{-t/{\tau}}$.  Should $\tau_{scint}$  be much smaller than all the phonon time constants, Eq.~\ref{distefano:eqscint} would simplify to Eq.~\ref{distefano:eqdirect} : $S = R$.
Fits have been performed independently on the direct hits using Eq.~\ref{distefano:eqdirect} 
and on the scintillation events using Eq.~\ref{distefano:eqscint}. 
For each detector, the common calorimetric parameters are found to be compatible  
within the error bars of both sets of fits.  
The resulting scintillation time constant is  
$\tau_{scint}= 0.4 \sys 0.1 \mbox{ ms}$,  
where the error quoted is systematic.  When the parameters of the direct fit are imposed on the scintillation fit, the result is degraded.  This indicates that there may in fact be several scintillation time constants.  The scintillation emission written in Eq.~\ref{distefano:eqemis} can be generalized to two or more scintillation constants, for instance $ H(t) \left[ \alpha \frac{{\mbox{e}}^{-t/{\tau_{scint}}}}{\tau_{scint}} + ( 1 - \alpha ) \frac{{\mbox{e}}^{-t/{\tau_{scint} ' }}}{\tau_{scint} '}\right] $.  Generalization of Eq.~\ref{distefano:eqscint} is linear.  With the direct-fit parameters set, fits yield a fast time constant of  $0.3 \sys 0.1$~ms making up $ \alpha = ( 70 \sys 15 )$~\% of the emitted scintillation energy, and a slow constant of $2.5 \sys 1$~ms.  More complicated emission time structures are possible but have not been investigated.

We note also that because the light detectors are not fully calorimetric for 
the scintillating light, pulse height is a biased 
estimator of the actual energy of the scintillation-induced pulses.  Compared 
to the energy scales provided by the direct hits (Eq.~\ref{distefano:eqdirect}), the energy of the 
scintillation pulses (Eq.~\ref{distefano:eqscint}) may be greatly underestimated.  
To correct for this, pulse parameters are obtained from the fit with a single scintillation time constant and all parameters left free.  These parameters are used to build a numerical pulse model based on Eq.~\ref{distefano:eqscint}.  The correction factor is defined as the ratio of 
the amplitude of this model pulse, assuming a scintillation time constant much shorter than the other time constants (i.e. $S_{\tau_{scint} \rightarrow 0} = R$, ideal calorimetric response),
divided by 
the amplitude of the model pulse with the fitted time constant (i.e. $S_{\tau_{scint}}$, actual response). 
Because each detector has its own phonon time constants, the correction factor varies from one detector to the next.  The correction factors are $1.45 \sys 0.15$ for the textured detector, $1.7 \sys 0.2$   for the small planar detector, and $1.2 \sys 0.1$ for the large planar detector.  The main errors are systematic and come from the estimation of the scintillation time.
Most of the loss in amplitude comes from the athermal component, because of its shorter time constant.  For the three detectors, the athermal correction factors are respectively $1.55 \sys 0.15$, $2.1 \sys 0.3$ and $1.4 \sys 0.1$, whereas they are only $1.17 \sys 0.05$, $1.45 \sys 0.1$ and $1.12 \sys 0.03$ for the slower thermal component.
One way to improve  the detector design might be to slow down the relaxation time of the 
athermal component, with a weaker coupling of the thermometer to the heat bath.

\subsection{Detector resolutions}
Resolutions for the direct \Fe\ hits and the scintillating \CoS\ hits are 
given in Table~\ref{distefano_tab:si_results_resolutions}. 
These \Fe\ resolutions are worse than the spot resolutions, 
confirming that the response of the light detector depends on the position of the interaction in it.  
This is because in the thin absorber crystal used, to reach the thermometer, phonons created far from it must bounce off the crystal surfaces and thus risk decay more often than phonons created near it.
Moreover, in the case of the non-textured silicon detectors, which have similar volume-to-surface ratios (0.25~mm), 
the resolutions  are close to one another (about 0.62~keV FWHM).  
The resolution of the textured detector, however, is noticeably worse.  
We attribute this to the lower volume-to-effective-surface ratio (0.15~mm) 
of the light detector, which causes extra scattering of the athermal phonons on the crystal surface before they are absorbed 
in the thin-film~\cite{art:proebst_1995}.

\CoS\ resolution is comparable in all three detectors.  Given 
 an energy of 
 $\approx 2.8 \mbox{ eV}$ 
 per scintillating photon, 
 this resolution is not compatible with simple Poisson photon statistics.  Rather, it appears to result from fluctuations in the light escaping the crystal depending on the position of the interaction in the scintillator because of flaws in the latter.  
 This effect may be more or less pronounced from one \CaW\ crystal to another.

Lastly, the $5 \sigma$ baseline noise, also reported in Tab.~\ref{distefano_tab:si_results_resolutions}, 
varies between 40~eV and 70~eV.
This gives an indication of detector threshold, though the high background from cosmic rays interacting in the scintillator 
has thwarted attempts  to verify it directly.

\subsection{Light absorption properties}
Thanks to the combined use of the \Fe\ and \CoS\ sources, 
it was possible to measure the absolute 
amount of light detected by each silicon calorimeter in this particular setup.
The energy deposited by a photon in the \CaW\ scintillator can go into three channels:
\begin{equation}
E_{dep} = E_{phonon} + E_{scint} + E_{lost}
\label{distefano:eqenergyscint}
\end{equation}
where $E_{phonon}$ is transferred to phonons in \CaW\ (either directly or by absorption of some scintillation photons).  $E_{scint}$ represents the energy of scintillation photons that are eventually seen in the light detector.  It can be measured absolutely thanks to the \Fe\ calibration of the light detector. $E_{lost}$ is energy lost, probably mainly in the form of scintillation photons escaping or absorbed in the light collector.  $E_{dep}$ itself is known in the case of a photon source like \CoS\  by identification of the photon lines.  This method does not provide a means of estimating $E_{phonon}$ and $E_{lost}$, though the former is thought  to represent the majority of the deposited energy.

Results for the three detectors are summarized in Table~\ref{distefano_tab:si_results} 
in terms of $\varepsilon = \frac{E_{scint}}{E_{dep}}$, the fraction of the energy  deposited by a photon 
in the scintillator that is converted to light and eventually seen in the light detector. 
The main uncertainty, discussed previously, is of a systematic nature and comes from the difficulty of estimating the true pulse energy because the response of the light detector is rather fast for the scintillating time structure.  A smaller systematic effect  comes 
from possible non-linearities in detector response due to irregular transition curves.   
An upper limit on this effect has been estimated using the film pulser 
in the case of the small detectors, and different operating points on all detectors.  
The effect was also reduced in the case of the large detector 
through the active thermal feedback~\cite{proc:OllieLTD8} for some runs.

The non-textured silicon detectors absorb comparable amounts of light per unit area.  
This indicates that detector size could still be increased before the quantity 
of absorbed light  reaches a limit for this combination of light 
collector and scintillator. 
The textured detector has a significantly better light absorption per unit area.  
Light absorption of the textured detector is 
in line with its effective surface area, though the estimation of ${\cal S}_{eff}$ is rough and assumes
that light falls uniformly on both faces of the detector, and that the small \SiN\ and W structures on the front face of the detector have reflectivities similar to Si.

Lastly, we note that the absolute amount of light seen in the light detector 
depends on the efficiency of the light collector, 
and also on the quality of the scintillator.  Indeed, significant variations 
have been observed between \CaW\ crystals leading to values of 
$\varepsilon$ of up to about 1.3~\% for configurations similar to those described here~\cite{proc:TorstenComo2001}.

\section{Conclusion}
The absorptivity in the visible wavelengths of cryogenic silicon calorimeters with a given size has been enhanced 
through the standard
photovoltaic technique of texturing.  
The increase is compatible with the extra surface created by the texturing (a factor of 1.74), 
as expected for photons with random incidence angles.  
We presume that for normal-incident photons, 
the gain would be greater due to photons having two chances 
to be absorbed between surface pyramids as is the case for solar cells.  

Response of the cryogenic light detector has been corrected for the mis-adaptation between its  time constants and the scintillation time structure of the \CaW\ crystal used as light source at 20~mK.  
The latter is estimated to contain at least a fast ($\approx 0.3$~ms) component and a slow ($\approx 2.5$~ms) one.  A slower light detector than the one used here would allow a better estimation of the scintillation energies. 
Conversely, a faster one might provide more insight into the time structure of the scintillator, and, at energies high enough to yield a statistically significant number of photons, could make it possible to check if the time structure depends on the nature of the interaction in the scintillator.  Indeed differences have already been observed between the time constants of alpha particles and those of gammas interacting in \CaW\ at 20~mK~\cite{thesis:Torsten2002}.

Energy resolution for a monochromatic source, which in the case 
of two-dimensional athermal silicon devices is dominated by the correlation 
between response and position of the interaction, 
is degraded by the decrease in volume-to-surface ratio caused by texturing, which induces extra phonon scattering.
It may be possible to improve the sensitivity of the light detector 
through the use of smaller thermometers with phonon collector 
pads~\cite{art:MartinNIMA2001}.
Another option might be to exploit the Neganov-Luke effect~\cite{art:Neganov_1993,art:Luke1990}, 
the semiconducting analogue of resistive heating, which can amplify the phonon signal.

Finally, for low-radioactivity applications, contaminations from the naturally occurring unstable isotope \Kf\ should be avoided.  The tungsten thermometer etch used here would therefore be replaced by a potassium-free one such as HF or $\mbox{H}_2\mbox{O}_2$.  It may also be possible to replace the KOH texturing bath with NaOH or an organic
one~\cite{art:TMAH}.

This
work was funded by European TMR Network for Cryogenic Detectors
ERB-FMRX-CT98-0167, and is dedicated to the memory of Marie Folkart.

\clearpage

{\Large List of references} \\

\clearpage

{\Large Tables} \\

\begin{table}[h]
\begin{center}
   \caption[Description of detectors]{Silicon light detectors.  ${\cal S}$ is the total surface area of both faces of the light detectors. Effective surface ${\cal S}_{eff}$  encompasses effect
   of texturing.  Volume is ${\cal V}$.  All devices were 0.5~mm thick.}
   \label{distefano_tab:si_description}
   \begin{tabular}{lrrrr} 
      \hline 
      \hline 
      Type & Size & ${\cal S}$ & ${\cal S}_{eff}$  & ${\cal V}$ \\
       &  ($\mbox{mm}^2$) &  ($\mbox{mm}^2$) &  ($\mbox{mm}^2$) & $ (\mbox{mm}^3)$ \\
	\hline
      Planar--Small & $20 \times 20$ & 800  & $800$ & 200 \\
      Planar--Big & $30 \times 30$ & 1800 & $1800$ & 450 \\
      Textured & $20 \times 20$ & $800$ & $1377$ & 200 \\
      \hline 
      \hline 
   \end{tabular} 
\end{center}
\end{table}

\clearpage

\begin{table}[h]
\begin{center}
   \caption[Resolutions of detectors]{Resolutions of detectors for direct \Fe\ and scintillating \CoS\ calibrations, and baseline noise.  
   \Fe\ resolution is mainly due to inhomogeneous response of light detector.  
   This is exacerbated in the case of the textured detector, which has a lower volume-to-effective-surface ratio.  
   \CoS\ resolutions, expressed in units of energy deposited in the 
   scintillator, are comparable for the three detectors, 
   and probably dominated by inhomogeneous response of the scintillator.  
   Baseline noise, an indicator of threshold, is also comparable in all detectors.}
   \label{distefano_tab:si_results_resolutions}
   \begin{tabular}{lrrrr} 
      \hline 
      \hline 
       Detector & $\mbox{FWHM}_{\mbox{Fe}}$ & $\mbox{FWHM}_{\mbox{Co}}$ & $5 \sigma_{\mbox{noise}}$ & $ \mbox{FWHM}_{\mbox{Fe}} \times {\cal V} / {\cal S}_{eff}$ \\		
	& (keV) & (MeV ee) & (eV) & $\mbox{ (keV\ mm)} $ \\
	\hline
      Planar--Small & $0.65$ & $0.26$ & $ 68 $ & 0.16 \\
      Planar--Big & $0.61$ & $0.30$ & $ 73 $ & 0.15 \\
      Textured & $1.25$ & $0.28$ & $ 39 $ & 0.18 \\
      \hline 
      \hline 
   \end{tabular} 
\end{center}
\end{table}

\clearpage

\begin{table}[h]
\begin{center}
   \caption[Light absorption of detectors]{Light absorption properties of Si light detectors. $\varepsilon$ is the fraction of energy deposited by \CoS\ photons in the scintillator and eventually detected in the light
   detector.  Non-textured silicon detectors exhibit comparable absorption per unit area.  This indicates that total light absorption could still be improved by a larger surface area in this particular setup.  Textured light detector has greater absorption per unit area.  Taking into account  effective area brings results into line, though estimating ${\cal S}_{eff}$ precisely is not straightforward.  Errors are systematic and come mainly from the estimate of the scintillation energy, based on pulse amplitude, which depends on knowledge of the scintillation time structure.}
   \label{distefano_tab:si_results}
   \begin{tabular}{lrrrr} 
      \hline 
      \hline 
       Detector & $10^{3} \times \varepsilon$ & $\varepsilon / {\cal S} $ & $\varepsilon / {\cal S}_{eff} $  \\
        &  & $ \mbox{ (m}^{-2})$ & $ \mbox{ (m}^{-2})$  \\
	\hline
      Planar--Small  & $\ 3.0 \pm 0.3 \ $ & $\ 3.8 \pm 0.4 \ $ & $\ 3.8 \pm 0.4 \ $ \\
      Planar--Big & $\ 7.8 \pm 0.8 \ $ & $\ 4.3 \pm 0.5 \ $ & $ 4.3 \pm 0.5 \ $ \\
      Textured & $\ 5.4 \pm 0.5 \ $ & $\ 6.7 \pm 0.7 \ $ & $ 3.9 \pm 0.4 \ $ \\
      \hline 
      \hline 
   \end{tabular} 
\end{center}
\end{table}

\clearpage

{\Large Figure captions} \\

\begin{figure}[h]
	\centering
	\caption[Geometry of thermometer]{Schematic of thermometers deposited on the Si cryogenic detectors.  
	Typical dimension of tungsten film is 2~mm.  Thick lines are Au wire, 
	used for thermal contacts and resistive heaters; thin lines are Al wire, 
	used as electrical contacts.}
	\label{distefano_fig:thermometer}
\end{figure}

\begin{figure}[h]
	\centering
	\caption[Overall scintillation setup]{Exploded view of copper holder, scintillator and light 
	detector (adapted from~\cite{proc:TorstenComo2001}).  Holder is lined with reflective foil, omitted here for 
	clarity.  Light detector is exposed to direct photon hits coming from 
	a \Fe\ source and from scintillation photons coming from the \CaW\ 
	excited by cosmic rays and a \CoS\ source.  The combination of direct 
	and scintillating events makes it possible to measure the absolute 
	amount of light detected for a given setup.}
	\label{distefano_fig:setup}
\end{figure}

\begin{figure}[h]
  \centering
  \caption[Direct and scintillation pulse shapes]{Examples of averaged direct and scintillation-induced pulses from a detector (continuous lines) along with their fits (dashed lines).  Pulses here have been averaged over some thirty events.  Fits have been performed leaving all parameters free in both cases and assume a single scintillation time constant.  The common, calorimetric parameters, are found to be compatible. When the fit parameters from the direct fit are imposed on the scintillation fit, the result is degraded.  This indicates there is in fact more than one scintillation time constant.}
  \label{fig:distefano_pulses}
\end{figure}

\clearpage

  \centering
	\mbox{\epsfig{file=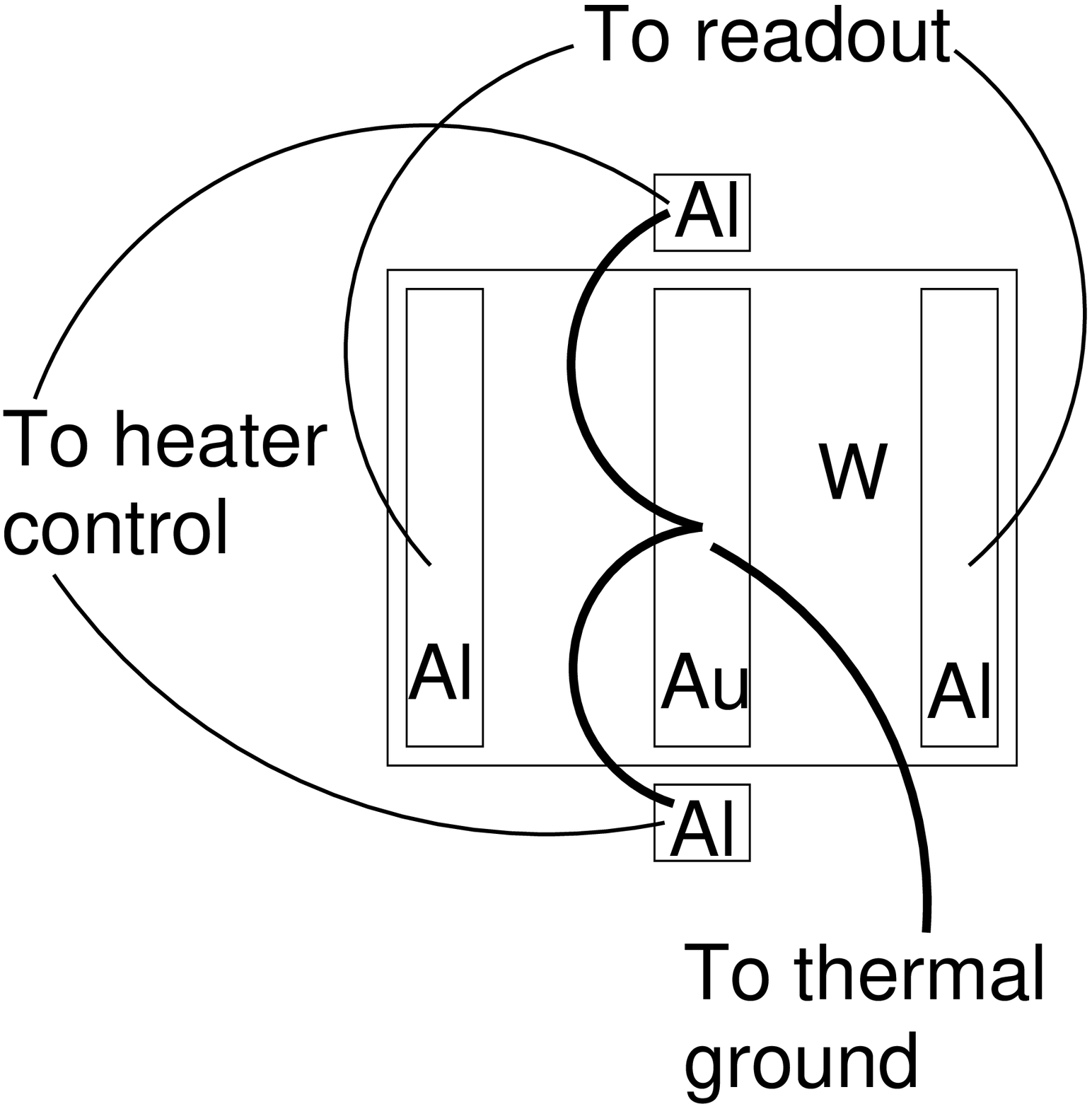,width=86mm}}

\clearpage

  \centering
	\mbox{\epsfig{file=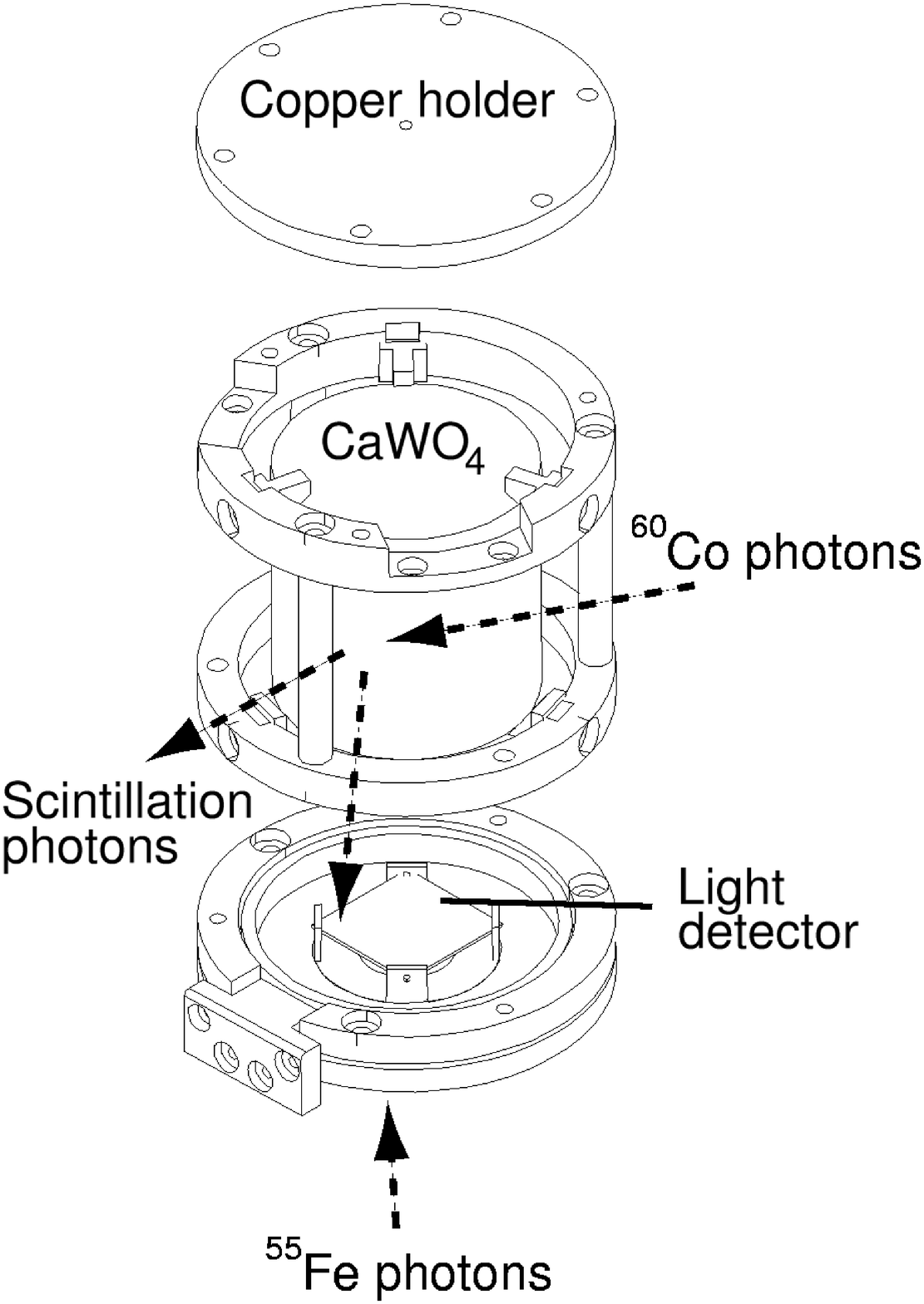,width=86mm}}

\clearpage

  \centering
        \mbox{\epsfig{file=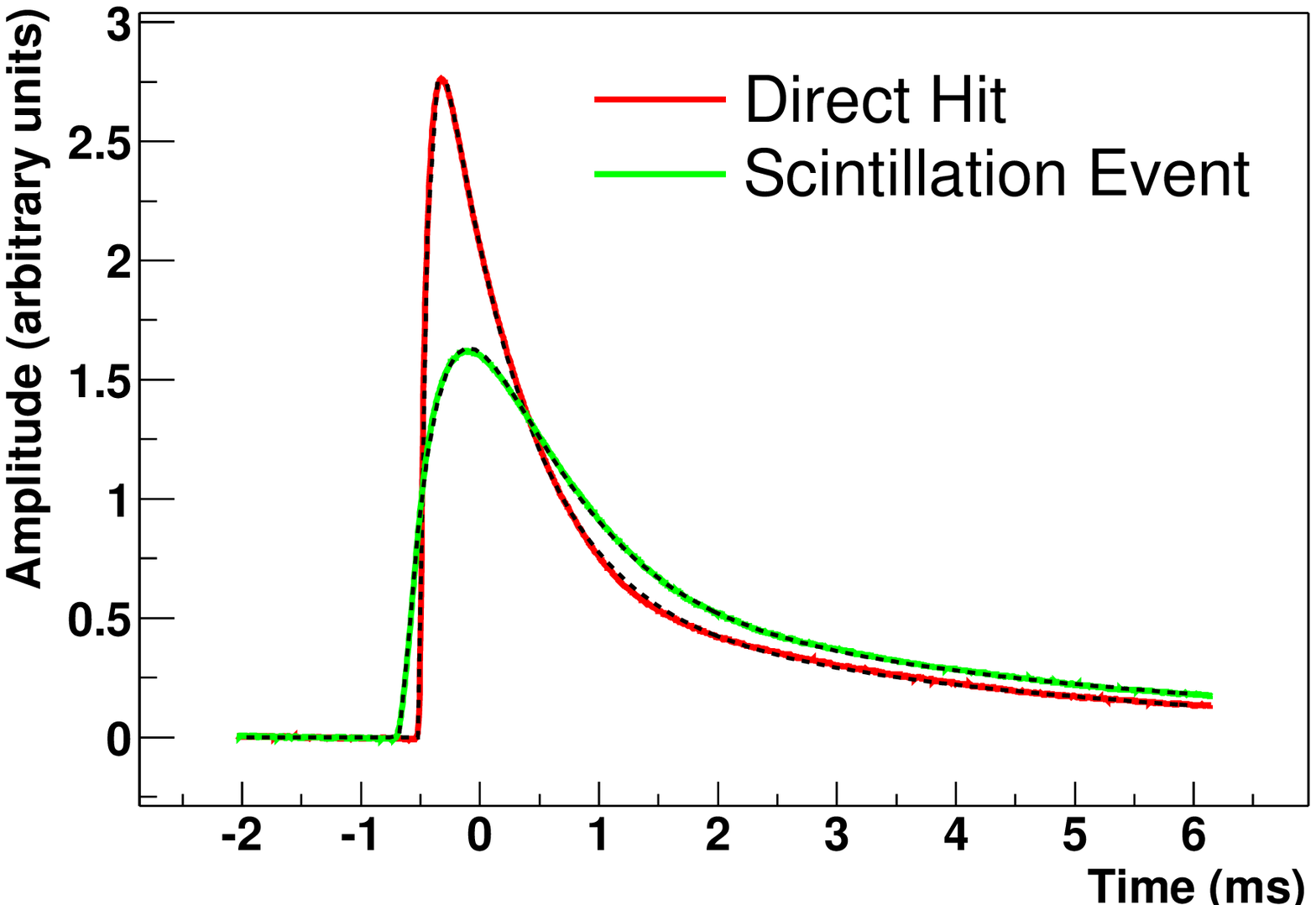,width=86mm}}


\begin{thebibliography}{20}
\expandafter\ifx\csname natexlab\endcsname\relax\def\natexlab#1{#1}\fi
\expandafter\ifx\csname bibnamefont\endcsname\relax
  \def\bibnamefont#1{#1}\fi
\expandafter\ifx\csname bibfnamefont\endcsname\relax
  \def\bibfnamefont#1{#1}\fi
\expandafter\ifx\csname citenamefont\endcsname\relax
  \def\citenamefont#1{#1}\fi
\expandafter\ifx\csname url\endcsname\relax
  \def\url#1{\texttt{#1}}\fi
\expandafter\ifx\csname urlprefix\endcsname\relax\def\urlprefix{URL }\fi
\providecommand{\bibinfo}[2]{#2}
\providecommand{\eprint}[2][]{\url{#2}}

\bibitem[{\citenamefont{Spooner et~al.}(1991)\citenamefont{Spooner, Bewick,
  Homer, Smith, and Lewin}}]{art:Spooner1991}
\bibinfo{author}{\bibfnamefont{N.~J.} \bibnamefont{Spooner}},
  \bibinfo{author}{\bibfnamefont{A.}~\bibnamefont{Bewick}},
  \bibinfo{author}{\bibfnamefont{G.}~\bibnamefont{Homer}},
  \bibinfo{author}{\bibfnamefont{P.}~\bibnamefont{Smith}}, \bibnamefont{and}
  \bibinfo{author}{\bibfnamefont{J.}~\bibnamefont{Lewin}},
  \bibinfo{journal}{Phys. Lett. B} \textbf{\bibinfo{volume}{273}},
  \bibinfo{pages}{333} (\bibinfo{year}{1991}).

\bibitem[{\citenamefont{Shutt et~al.}(1992)\citenamefont{Shutt, Ellman, Barnes,
  Cummings, Da~Silva, Emes, Giraud-H\'eraud, Haller, Lange, Ross
  et~al.}}]{art:Shutt}
\bibinfo{author}{\bibfnamefont{T.}~\bibnamefont{Shutt}},
  \bibinfo{author}{\bibfnamefont{B.}~\bibnamefont{Ellman}},
  \bibinfo{author}{\bibfnamefont{P.~D.} \bibnamefont{Barnes},
  \bibfnamefont{Jr.}},
  \bibinfo{author}{\bibfnamefont{A.}~\bibnamefont{Cummings}},
  \bibinfo{author}{\bibfnamefont{A.}~\bibnamefont{Da~Silva}},
  \bibinfo{author}{\bibfnamefont{J.}~\bibnamefont{Emes}},
  \bibinfo{author}{\bibfnamefont{Y.}~\bibnamefont{Giraud-H\'eraud}},
  \bibinfo{author}{\bibfnamefont{E.~E.} \bibnamefont{Haller}},
  \bibinfo{author}{\bibfnamefont{A.~E.} \bibnamefont{Lange}},
  \bibinfo{author}{\bibfnamefont{R.~R.} \bibnamefont{Ross}},
  \bibnamefont{et~al.}, \bibinfo{journal}{Phys. Rev. Lett.}
  \textbf{\bibinfo{volume}{69}}, \bibinfo{pages}{3425} (\bibinfo{year}{1992}).

\bibitem[{\citenamefont{Berg\'e et~al.}(1999)\citenamefont{Berg\'e, Berkes,
  Chambon, Chapellier, Chardin, Charvin, De~J\'esus, Di~Stefano, Drain,
  Dumoulin et~al.}}]{proc:Edelweiss2}
\bibinfo{author}{\bibfnamefont{L.}~\bibnamefont{Berg\'e}},
  \bibinfo{author}{\bibfnamefont{I.}~\bibnamefont{Berkes}},
  \bibinfo{author}{\bibfnamefont{B.}~\bibnamefont{Chambon}},
  \bibinfo{author}{\bibfnamefont{M.}~\bibnamefont{Chapellier}},
  \bibinfo{author}{\bibfnamefont{G.}~\bibnamefont{Chardin}},
  \bibinfo{author}{\bibfnamefont{P.}~\bibnamefont{Charvin}},
  \bibinfo{author}{\bibfnamefont{M.}~\bibnamefont{De~J\'esus}},
  \bibinfo{author}{\bibfnamefont{P.}~\bibnamefont{Di~Stefano}},
  \bibinfo{author}{\bibfnamefont{D.}~\bibnamefont{Drain}},
  \bibinfo{author}{\bibfnamefont{L.}~\bibnamefont{Dumoulin}},
  \bibnamefont{et~al.}, \bibinfo{journal}{Nucl. Phys. B (Proc. Suppl.)}
  \textbf{\bibinfo{volume}{70}}, \bibinfo{pages}{69} (\bibinfo{year}{1999}),
  \bibinfo{note}{{astro-ph/}9801199}.

\bibitem[{\citenamefont{Bobin et~al.}(1997)\citenamefont{Bobin, Berkes,
  Hadjout, Coron, Leblanc, and de~Marcillac}}]{art:Bobin_Scint1997}
\bibinfo{author}{\bibfnamefont{C.}~\bibnamefont{Bobin}},
  \bibinfo{author}{\bibfnamefont{I.}~\bibnamefont{Berkes}},
  \bibinfo{author}{\bibfnamefont{J.~P.} \bibnamefont{Hadjout}},
  \bibinfo{author}{\bibfnamefont{N.}~\bibnamefont{Coron}},
  \bibinfo{author}{\bibfnamefont{J.}~\bibnamefont{Leblanc}}, \bibnamefont{and}
  \bibinfo{author}{\bibfnamefont{P.}~\bibnamefont{de~Marcillac}},
  \bibinfo{journal}{Nucl. Instrum. Methods A} \textbf{\bibinfo{volume}{386}},
  \bibinfo{pages}{453} (\bibinfo{year}{1997}).

\bibitem[{\citenamefont{Meunier et~al.}(1999)\citenamefont{Meunier, Bravin,
  Bruckmayer, Giordano, Loidl, Meier, Pr\"obst, Seidel, Sisti, Stodolsky
  et~al.}}]{art:Meunier1999}
\bibinfo{author}{\bibfnamefont{P.}~\bibnamefont{Meunier}},
  \bibinfo{author}{\bibfnamefont{M.}~\bibnamefont{Bravin}},
  \bibinfo{author}{\bibfnamefont{M.}~\bibnamefont{Bruckmayer}},
  \bibinfo{author}{\bibfnamefont{S.}~\bibnamefont{Giordano}},
  \bibinfo{author}{\bibfnamefont{M.}~\bibnamefont{Loidl}},
  \bibinfo{author}{\bibfnamefont{O.}~\bibnamefont{Meier}},
  \bibinfo{author}{\bibfnamefont{F.}~\bibnamefont{Pr\"obst}},
  \bibinfo{author}{\bibfnamefont{W.}~\bibnamefont{Seidel}},
  \bibinfo{author}{\bibfnamefont{M.}~\bibnamefont{Sisti}},
  \bibinfo{author}{\bibfnamefont{L.}~\bibnamefont{Stodolsky}},
  \bibnamefont{et~al.}, \bibinfo{journal}{Appl. Phys. Lett.}
  \textbf{\bibinfo{volume}{75}}, \bibinfo{pages}{1335} (\bibinfo{year}{1999}),
  \bibinfo{note}{{physics/}9906017}.

\bibitem[{\citenamefont{Alessandrello et~al.}(1998)\citenamefont{Alessandrello,
  Bashkirov, Brofferio, Bucci, Camin, Cremonesi, Fiorini, Gervasio, Giuliani,
  Nucciotti et~al.}}]{art:Allesandrello_Scint1998}
\bibinfo{author}{\bibfnamefont{A.}~\bibnamefont{Alessandrello}},
  \bibinfo{author}{\bibfnamefont{V.}~\bibnamefont{Bashkirov}},
  \bibinfo{author}{\bibfnamefont{C.}~\bibnamefont{Brofferio}},
  \bibinfo{author}{\bibfnamefont{C.}~\bibnamefont{Bucci}},
  \bibinfo{author}{\bibfnamefont{D.}~\bibnamefont{Camin}},
  \bibinfo{author}{\bibfnamefont{O.}~\bibnamefont{Cremonesi}},
  \bibinfo{author}{\bibfnamefont{E.}~\bibnamefont{Fiorini}},
  \bibinfo{author}{\bibfnamefont{G.}~\bibnamefont{Gervasio}},
  \bibinfo{author}{\bibfnamefont{A.}~\bibnamefont{Giuliani}},
  \bibinfo{author}{\bibfnamefont{A.}~\bibnamefont{Nucciotti}},
  \bibnamefont{et~al.}, \bibinfo{journal}{Phys. Lett. B}
  \textbf{\bibinfo{volume}{420}}, \bibinfo{pages}{109} (\bibinfo{year}{1998}).

\bibitem[{\citenamefont{Bucci et~al.}(2001)\citenamefont{Bucci, Altmann,
  Bruckmayer, Cozzini, Di~Stefano, Frank, Hauff, Pr\"obst, Seidel, Sergeyev
  et~al.}}]{preprint:CRESST_proposal_2001}
\bibinfo{author}{\bibfnamefont{C.}~\bibnamefont{Bucci}},
  \bibinfo{author}{\bibfnamefont{M.}~\bibnamefont{Altmann}},
  \bibinfo{author}{\bibfnamefont{M.}~\bibnamefont{Bruckmayer}},
  \bibinfo{author}{\bibfnamefont{C.}~\bibnamefont{Cozzini}},
  \bibinfo{author}{\bibfnamefont{P.}~\bibnamefont{Di~Stefano}},
  \bibinfo{author}{\bibfnamefont{T.}~\bibnamefont{Frank}},
  \bibinfo{author}{\bibfnamefont{D.}~\bibnamefont{Hauff}},
  \bibinfo{author}{\bibfnamefont{F.}~\bibnamefont{Pr\"obst}},
  \bibinfo{author}{\bibfnamefont{W.}~\bibnamefont{Seidel}},
  \bibinfo{author}{\bibfnamefont{I.}~\bibnamefont{Sergeyev}},
  \bibnamefont{et~al.} (\bibinfo{year}{2001}),
  \bibinfo{note}{{MPI-PhE{/}2001-02}}.

\bibitem[{\citenamefont{Pr\"obst et~al.}(1995)\citenamefont{Pr\"obst, Frank,
  Cooper, Colling, Dummer, Ferger, Forster, Nucciotti, Seidel, and
  Stodolsky}}]{art:proebst_1995}
\bibinfo{author}{\bibfnamefont{F.}~\bibnamefont{Pr\"obst}},
  \bibinfo{author}{\bibfnamefont{M.}~\bibnamefont{Frank}},
  \bibinfo{author}{\bibfnamefont{S.}~\bibnamefont{Cooper}},
  \bibinfo{author}{\bibfnamefont{P.}~\bibnamefont{Colling}},
  \bibinfo{author}{\bibfnamefont{D.}~\bibnamefont{Dummer}},
  \bibinfo{author}{\bibfnamefont{P.}~\bibnamefont{Ferger}},
  \bibinfo{author}{\bibfnamefont{G.}~\bibnamefont{Forster}},
  \bibinfo{author}{\bibfnamefont{A.}~\bibnamefont{Nucciotti}},
  \bibinfo{author}{\bibfnamefont{W.}~\bibnamefont{Seidel}}, \bibnamefont{and}
  \bibinfo{author}{\bibfnamefont{L.}~\bibnamefont{Stodolsky}},
  \bibinfo{journal}{J. Low Temp. Phys.} \textbf{\bibinfo{volume}{100}},
  \bibinfo{pages}{69} (\bibinfo{year}{1995}).

\bibitem[{\citenamefont{Arndt et~al.}(1975)\citenamefont{Arndt, Allison,
  Haynos, and Meulenberg}}]{proc:arndt1975}
\bibinfo{author}{\bibfnamefont{R.~A.} \bibnamefont{Arndt}},
  \bibinfo{author}{\bibfnamefont{J.~F.} \bibnamefont{Allison}},
  \bibinfo{author}{\bibfnamefont{J.~G.} \bibnamefont{Haynos}},
  \bibnamefont{and}
  \bibinfo{author}{\bibfnamefont{A.}~\bibnamefont{Meulenberg},
  \bibfnamefont{Jr.}}, in \emph{\bibinfo{booktitle}{11th {IEEE} Photovoltaic
  Specialists Conf. ({P}hoenix, {A}rizona)}} (\bibinfo{year}{1975}), pp.
  \bibinfo{pages}{40--43}.

\bibitem[{\citenamefont{Price}(1973)}]{proc:IPAetching1973}
\bibinfo{author}{\bibfnamefont{J.~B.} \bibnamefont{Price}}, in
  \emph{\bibinfo{booktitle}{Semiconductor Silicon Symposium}}
  (\bibinfo{year}{1973}), The Electrochemical Society Series, pp.
  \bibinfo{pages}{339--353}.

\bibitem[{\citenamefont{Colling et~al.}(1995)\citenamefont{Colling, Nucciotti,
  Bucci, Cooper, Ferger, Frank, Nagel, Pr\"obst, and Seidel}}]{art:Colling1995}
\bibinfo{author}{\bibfnamefont{P.}~\bibnamefont{Colling}},
  \bibinfo{author}{\bibfnamefont{A.}~\bibnamefont{Nucciotti}},
  \bibinfo{author}{\bibfnamefont{C.}~\bibnamefont{Bucci}},
  \bibinfo{author}{\bibfnamefont{S.}~\bibnamefont{Cooper}},
  \bibinfo{author}{\bibfnamefont{P.}~\bibnamefont{Ferger}},
  \bibinfo{author}{\bibfnamefont{M.}~\bibnamefont{Frank}},
  \bibinfo{author}{\bibfnamefont{U.}~\bibnamefont{Nagel}},
  \bibinfo{author}{\bibfnamefont{F.}~\bibnamefont{Pr\"obst}}, \bibnamefont{and}
  \bibinfo{author}{\bibfnamefont{W.}~\bibnamefont{Seidel}},
  \bibinfo{journal}{Nucl. Instrum. Methods A} \textbf{\bibinfo{volume}{354}},
  \bibinfo{pages}{408} (\bibinfo{year}{1995}).

\bibitem[{\citenamefont{Frank et~al.}(2001)\citenamefont{Frank, Bruckmayer,
  Cozzini, Di~Stefano, Hauff, Pr\"obst, Seidel, Angloher, and
  Schmidt}}]{proc:TorstenComo2001}
\bibinfo{author}{\bibfnamefont{T.}~\bibnamefont{Frank}},
  \bibinfo{author}{\bibfnamefont{M.}~\bibnamefont{Bruckmayer}},
  \bibinfo{author}{\bibfnamefont{C.}~\bibnamefont{Cozzini}},
  \bibinfo{author}{\bibfnamefont{P.}~\bibnamefont{Di~Stefano}},
  \bibinfo{author}{\bibfnamefont{D.}~\bibnamefont{Hauff}},
  \bibinfo{author}{\bibfnamefont{F.}~\bibnamefont{Pr\"obst}},
  \bibinfo{author}{\bibfnamefont{W.}~\bibnamefont{Seidel}},
  \bibinfo{author}{\bibfnamefont{G.}~\bibnamefont{Angloher}}, \bibnamefont{and}
  \bibinfo{author}{\bibfnamefont{J.}~\bibnamefont{Schmidt}}, in
  \emph{\bibinfo{booktitle}{7th Int. Conf. Advanced Tech. and Part. Phys.
  ({V}illa {O}lmo, {C}omo, {I}taly)}} (\bibinfo{year}{2001}).

\bibitem[{\citenamefont{Frank}(2002)}]{thesis:Torsten2002}
\bibinfo{author}{\bibfnamefont{T.}~\bibnamefont{Frank}}, Ph.D. thesis,
  \bibinfo{school}{Technische Universit\"at M\"unchen and Max-Planck-Institut
  f\"ur Physik, Werner-Heisenberg-Institut, Munich} (\bibinfo{year}{2002}),
  \bibinfo{note}{http://tumb1.biblio.tu-muenchen.de/publ/diss/ph/2002/frank.pd%
f}.

\bibitem[{\citenamefont{Weber et~al.}(2000)\citenamefont{Weber, Stover,
  Gilbert, Nevitt, and Ouderkirk}}]{art:PolyFoil}
\bibinfo{author}{\bibfnamefont{M.~F.} \bibnamefont{Weber}},
  \bibinfo{author}{\bibfnamefont{C.~A.} \bibnamefont{Stover}},
  \bibinfo{author}{\bibfnamefont{L.~R.} \bibnamefont{Gilbert}},
  \bibinfo{author}{\bibfnamefont{T.~J.} \bibnamefont{Nevitt}},
  \bibnamefont{and} \bibinfo{author}{\bibfnamefont{A.~J.}
  \bibnamefont{Ouderkirk}}, \bibinfo{journal}{Science}
  \textbf{\bibinfo{volume}{287}}, \bibinfo{pages}{2451} (\bibinfo{year}{2000}).

\bibitem[{\citenamefont{Treadaway and Powell}(1974)}]{art:Treadaway1974}
\bibinfo{author}{\bibfnamefont{M.~J.} \bibnamefont{Treadaway}}
  \bibnamefont{and} \bibinfo{author}{\bibfnamefont{R.~C.}
  \bibnamefont{Powell}}, \bibinfo{journal}{J. Chem. Phys.}
  \textbf{\bibinfo{volume}{61}}, \bibinfo{pages}{4003} (\bibinfo{year}{1974}).

\bibitem[{\citenamefont{Meier et~al.}(2000)\citenamefont{Meier, Bravin,
  Bruckmayer, Di~Stefano, Frank, Loidl, Meunier, Pr\"obst, Safran, Seidel
  et~al.}}]{proc:OllieLTD8}
\bibinfo{author}{\bibfnamefont{O.}~\bibnamefont{Meier}},
  \bibinfo{author}{\bibfnamefont{M.}~\bibnamefont{Bravin}},
  \bibinfo{author}{\bibfnamefont{M.}~\bibnamefont{Bruckmayer}},
  \bibinfo{author}{\bibfnamefont{P.}~\bibnamefont{Di~Stefano}},
  \bibinfo{author}{\bibfnamefont{T.}~\bibnamefont{Frank}},
  \bibinfo{author}{\bibfnamefont{M.}~\bibnamefont{Loidl}},
  \bibinfo{author}{\bibfnamefont{P.}~\bibnamefont{Meunier}},
  \bibinfo{author}{\bibfnamefont{F.}~\bibnamefont{Pr\"obst}},
  \bibinfo{author}{\bibfnamefont{G.}~\bibnamefont{Safran}},
  \bibinfo{author}{\bibfnamefont{W.}~\bibnamefont{Seidel}},
  \bibnamefont{et~al.}, \bibinfo{journal}{Nucl. Instrum. Methods A}
  \textbf{\bibinfo{volume}{444}}, \bibinfo{pages}{350} (\bibinfo{year}{2000}).

\bibitem[{\citenamefont{Loidl et~al.}(2001)\citenamefont{Loidl, Cooper, Meier,
  Pr\"obst, S\'afr\'an, Seidel, Sisti, Stodolsky, and
  Uchaikin}}]{art:MartinNIMA2001}
\bibinfo{author}{\bibfnamefont{M.}~\bibnamefont{Loidl}},
  \bibinfo{author}{\bibfnamefont{S.}~\bibnamefont{Cooper}},
  \bibinfo{author}{\bibfnamefont{O.}~\bibnamefont{Meier}},
  \bibinfo{author}{\bibfnamefont{F.}~\bibnamefont{Pr\"obst}},
  \bibinfo{author}{\bibfnamefont{G.}~\bibnamefont{S\'afr\'an}},
  \bibinfo{author}{\bibfnamefont{W.}~\bibnamefont{Seidel}},
  \bibinfo{author}{\bibfnamefont{M.}~\bibnamefont{Sisti}},
  \bibinfo{author}{\bibfnamefont{L.}~\bibnamefont{Stodolsky}},
  \bibnamefont{and} \bibinfo{author}{\bibfnamefont{S.}~\bibnamefont{Uchaikin}},
  \bibinfo{journal}{Nucl. Instrum. Methods A} \textbf{\bibinfo{volume}{465}},
  \bibinfo{pages}{440} (\bibinfo{year}{2001}).

\bibitem[{\citenamefont{Neganov et~al.}(1993)\citenamefont{Neganov, Trofimov,
  and Stepankin}}]{art:Neganov_1993}
\bibinfo{author}{\bibfnamefont{B.~S.} \bibnamefont{Neganov}},
  \bibinfo{author}{\bibfnamefont{V.~N.} \bibnamefont{Trofimov}},
  \bibnamefont{and}
  \bibinfo{author}{\bibfnamefont{V.}~\bibnamefont{Stepankin}},
  \bibinfo{journal}{J. Low Temp. Phys.} \textbf{\bibinfo{volume}{93}},
  \bibinfo{pages}{417} (\bibinfo{year}{1993}).

\bibitem[{\citenamefont{Luke et~al.}(1990)\citenamefont{Luke, Beeman, Goulding,
  Labov, and Silver}}]{art:Luke1990}
\bibinfo{author}{\bibfnamefont{P.}~\bibnamefont{Luke}},
  \bibinfo{author}{\bibfnamefont{J.}~\bibnamefont{Beeman}},
  \bibinfo{author}{\bibfnamefont{F.~S.} \bibnamefont{Goulding}},
  \bibinfo{author}{\bibfnamefont{S.~E.} \bibnamefont{Labov}}, \bibnamefont{and}
  \bibinfo{author}{\bibfnamefont{E.~H.} \bibnamefont{Silver}},
  \bibinfo{journal}{Nucl. Instrum. Methods A} \textbf{\bibinfo{volume}{289}},
  \bibinfo{pages}{406} (\bibinfo{year}{1990}).

\bibitem[{\citenamefont{You et~al.}(2001)\citenamefont{You, Kim, Huh, Park,
  Pak, and Kan}}]{art:TMAH}
\bibinfo{author}{\bibfnamefont{J.~S.} \bibnamefont{You}},
  \bibinfo{author}{\bibfnamefont{D.}~\bibnamefont{Kim}},
  \bibinfo{author}{\bibfnamefont{J.~Y.} \bibnamefont{Huh}},
  \bibinfo{author}{\bibfnamefont{H.~J.} \bibnamefont{Park}},
  \bibinfo{author}{\bibfnamefont{J.~J.} \bibnamefont{Pak}}, \bibnamefont{and}
  \bibinfo{author}{\bibfnamefont{C.~S.} \bibnamefont{Kan}},
  \bibinfo{journal}{Sol. En. Mat. \& Sol. Cells} \textbf{\bibinfo{volume}{66}},
  \bibinfo{pages}{37} (\bibinfo{year}{2001}).

\end{thebibliography}
\end{document}